\documentclass[oneside,english]{amsart}
\usepackage[T1]{fontenc}
\usepackage[latin9]{inputenc}
\usepackage{geometry}
\usepackage{amstext}
\usepackage{amsthm}
\usepackage{amssymb}

\makeatletter

\providecommand{\tabularnewline}{\\}

\numberwithin{equation}{section}
\numberwithin{figure}{section}
  \theoremstyle{remark}
  \newtheorem{rem}{\protect\remarkname}
  \theoremstyle{plain}
  \newtheorem{prop}{\protect\propositionname}

\makeatother

\usepackage{babel}
  \providecommand{\propositionname}{Proposition}
  \providecommand{\remarkname}{Remark}

\begin{document}

\title{The Co-terminal Swap Market Model with Bergomi Stochastic Volatility}

\author{Kenjiro Oya}
\begin{abstract}
In this article, we apply the forward variance modeling approach by
L.Bergomi to the co-terminal swap market model. We build an interest
rate model for which all the market price changes of hedging instruments,
interest rate swaps and European swaptions, are interpreted as the
state variable variations, and no diffusion parameter calibration
procedure is required. The model provides quite simple profit and
loss (PnL) formula, with which we can easily understand where a material
PnL trend comes from when it appears, and consider how we should modify
the model parameters. The model has high flexibility to control the
model dynamics because parameter calibration is unnecessary and the
model parameters can be used solely for the purpose of the model dynamics
control. With the model, the position management of the exotic interest
rate products, e.g. Bermudan swaptions, can be carried out in a more
sophisticated and systematic manner. A numerical experiment is performed
to show the effectiveness of the approach for a Canary swaption, which
is a special form of a Bermudan swaption.
\end{abstract}

\maketitle

\section{Introduction}

It is common practice to hedge volatility exposure of exotic derivative
products with vanilla options. However, the methodology of the volatility
exposure hedge is not fully established yet. A difficulty lies in
the fact that the direct modeling of vanilla option prices or implied
volatilities is technically challenging \cite{bergomi2015stochastic}.
Therefore, practitioners often deal with the issue by calibration
of diffusion parameters. Namely, diffusion parameters, which are assumed
to be constant in the model dynamics, are adjusted on a regular basis
so that the model could reproduce the market prices of vanilla options.
On the other hand, the diffusion parameters calibrated at different
timings will be inconsistent among them. Consequently, in a profit
and loss (PnL) analysis of a derivative contract, the PnL will contain
an additional contribution from the diffusion parameters change, which
is considered to be difficult to manage. In particular, when the PnL
results have an unexpected trend, it is not straightforward to know
how we should modify the model assumptions. 

In equity modeling, a promising approach to deal with the issue is
the forward variance model introduced by L.Bergomi in \cite{bergomi2005smile},
for which a forward variance curve is considered to be a model state
variable. For this model, all the market rate changes are understood
as the model state variable variations and the calibration of model
parameters are not required. This feature makes the PnL formula in
terms of the market observables quite simple, and the risk management
of the derivative contract more comfortable. However, in interest
rate modeling, the equivalent approach has not been presented to the
best knowledge of the author of the article. This might be because
there is no liquidity in variance swaps of interest rates. In reality,
the forward variance model can be still useful if forward variance
curves are computed using the market prices of vanilla options \cite{bergomi2015stochastic}.
In this article, we build a forward variance model for the co-terminal
swap market model such that 1) all the market price changes of hedging
instruments are interpreted as the state variable changes, 2) the
model has flexible parameters which are solely used for the control
of the model dynamics, and 3) the PnL formula becomes quite simple
so we can easily understand where a material PnL trend comes when
it appears, and consider how we should modify the model parameters.
With the model, the position management of the exotic interest rate
products, e.g. Bermudan swaptions, can be performed in a more sophisticated
and systematic manner.

The rest of the article is organized as follows. First, we review
the forward variance model. Next, we discuss the application of the
forward variance modeling approach to the swap market model. Then,
we present the co-terminal swap market Bergomi model (SMBM) and discuss
how we compute the state variables using market prices of European
swaptions. Lastly, we perform a numerical experiment using the co-terminal
SMBM.

\section{The Swap Market Model With Bergomi Stochastic Volatility}

In this section, we firstly review the forward variance model (Bergomi
stochastic volatility model) introduced in \cite{bergomi2005smile}.
Next, we consider a variance swap contract on a swap rate, with which
we discuss how we apply the forward variance modeling approach to
the swap market model. 

\subsection{A Forward Variance Curve and Bergomi Stochastic Volatility Model}

Bergomi stochastic volatility model uses a forward variance curve
as the modeling object. A variance swap is a contract that pays the
realized variance of the log-return of a tradable asset $S$ less
a strike rate $\varphi_{T}(t)$ at maturity $T$. We assume that the
payoff at $T$ is given as:
\begin{equation}
\frac{1}{T-t}\int_{t}^{T}\left(d\log S_{u}\right)^{2}-\varphi_{T}(t)\label{eq:VSDef}
\end{equation}

where $\varphi_{T}(t)$ is set at $t$ so as to make the value of
the variance swap zero. Namely, $\varphi_{T}(t)$ must satisfy:
\begin{eqnarray}
 &  & B_{t,T}^{-1}\left(\frac{1}{T-t}E_{t}^{Q}[\int_{t}^{T}\left(d\log S_{u}\right)^{2}]-\varphi_{T}(t)\right)\\
 & \Longleftrightarrow & \varphi_{T}(t)=\frac{1}{T-t}E_{t}^{Q}[\int_{t}^{T}\left(d\log S_{u}\right)^{2}]\label{eq:VarianceExpectation}
\end{eqnarray}

with a bank account process $B_{t,T}=\exp(\int_{t}^{T}r_{u}du)$ where
$r_{t}$ is a risk-free rate. $E_{t}^{Q}$ is the $t$-conditional
expectation under the risk-neutral measure. In the section 2.1, we
assume $r_{t}$ to be a deterministic process. 

A discrete forward variance swap rate $\varphi_{T_{1},T_{2}}(t)$
is defined using $\varphi_{T}(t)$ as:
\begin{equation}
\varphi_{T_{1},T_{2}}(t)\triangleq\frac{(T_{2}-t)\varphi_{T_{2}}(t)-(T_{1}-t)\varphi_{T_{1}}(t)}{T_{2}-T_{1}}
\end{equation}
With (\ref{eq:VarianceExpectation}), we can confirm that $\varphi_{T_{1},T_{2}}(t)$
is a martingale under the risk-neutral measure. For $T_{-}<T_{+}$,
we obtain:
\begin{equation}
E_{T_{-}}^{Q}[\varphi_{T_{1},T_{2}}(T_{+})]=E_{T_{-}}^{Q}[\frac{E_{T_{+}}^{Q}[\int_{T_{1}}^{T_{2}}\left(d\log S_{u}\right)^{2}]}{T_{2}-T_{1}}]=E_{T_{-}}^{Q}[\frac{\int_{T_{1}}^{T_{2}}\left(d\log S_{u}\right)^{2}}{T_{2}-T_{1}}]=\varphi_{T_{1},T_{2}}(T_{-})\label{eq:MartingaleConditionFwdVariance}
\end{equation}

An infinitesimal forward variance swap rate is obtained by taking
limit $\varepsilon\rightarrow0$ for $\varphi_{T,T+\varepsilon}(t)$:
\begin{equation}
\xi_{t}^{T}\triangleq\lim_{\varepsilon\rightarrow0}\varphi_{T,T+\varepsilon}(t)\label{eq:InfinitesimalVS}
\end{equation}

A continuous Bergomi model is specified by assuming log-normal dynamics
for infinitesimal forward variance swap rates in a forward variance
curve $\{\xi_{t}^{u}\}_{t<u\leq T_{e}}$:
\begin{equation}
d\xi_{t}^{u}=\omega_{t}^{u}\xi_{t}^{u}dW_{t}^{u},\;t<u\leq T_{e}
\end{equation}

where $\omega_{t}^{u}$ is a static model parameter, $W_{t}^{u}$
is a Brownian motion under the risk-neutral measure and $T_{e}$ is
the model terminal date. Note that $\xi_{t}^{T}$ has zero risk-neutral
drift because the forward variance swap rate is a martingale under
the risk-neutral measure as confirmed in (\ref{eq:MartingaleConditionFwdVariance}).
The underlying process $S_{t}$ follows:
\begin{equation}
dS_{t}=r_{t}S_{t}dt+\sigma^{S}S_{t}dW_{t}^{S}\label{eq:AssetSDEBergomi}
\end{equation}

where $W_{t}^{S}$ is a Brownian motion under the risk-neutral measure.
With (\ref{eq:VSDef}),(\ref{eq:InfinitesimalVS}) and (\ref{eq:AssetSDEBergomi}),
we obtain:
\begin{equation}
\xi_{t}^{t}=(d\log S_{t})^{2}/dt=\left(\sigma^{S}\right)^{2}
\end{equation}

Thus we get $\sigma^{S}=\sqrt{\xi_{t}^{t}}$. 

Bergomi model assumes $S_{t}$ and $\{\xi_{t}^{u}\}_{t\leq u\leq T_{e}}$
to be the model state variables. Consider that we are managing a derivative
contract using the underlying asset $S_{t}$ and the infinitesimal
forward variance swaps on $\{\xi_{t}^{u}\}_{t<u\leq T_{e}}$ as the
hedging instruments. We denote the value of the derivative contract
by $V(S_{t},\{\xi_{t}^{u}\}_{t\leq u\leq T_{e}},t)$. The pricing
equation is written as:
\begin{equation}
\frac{\partial V}{\partial t}-r_{t}V+\frac{\partial V}{\partial S}r_{t}S_{t}+\frac{1}{2}\frac{\partial^{2}V}{\partial S^{2}}\xi_{t}^{t}S_{t}^{2}+\int_{t}^{T_{e}}du\frac{\partial^{2}V}{\partial S\partial\xi^{u}}\rho_{t}^{u}\omega_{t}^{u}\xi_{t}^{u}\sqrt{\xi_{t}^{t}}S_{t}+\frac{1}{2}\int_{t}^{T_{e}}du\int_{t}^{T_{e}}dv\frac{\partial^{2}V}{\partial\xi^{u}\partial\xi^{v}}\rho_{t}^{uv}\omega_{t}^{u}\omega_{t}^{v}\xi_{t}^{u}\xi_{t}^{v}=0\label{eq:PricingEquationBergomi}
\end{equation}

with the correlation functions, $\langle dW^{u},dW^{S}\rangle_{t}=\rho_{t}^{u}dt$
and $\langle dW^{u},dW^{v}\rangle_{t}=\rho_{t}^{uv}dt$. 

Consider a hedged contract $V_{t}^{H}$ which is a portfolio of an
unit of the derivative contract $V_{t}$, a bank account $B_{0,t}$,
the underlying asset $S_{t}$ and the forward variance swaps on $\{\xi_{t}^{u}\}_{t<u\leq T_{e}}$.
We require $V_{t}^{H}$ to satisfy $\partial_{S}V_{t}^{H}=0,\;\partial_{\xi^{u}}V_{t}^{H}=0,\;t<u\leq T_{e}$
and $V_{t}^{H}=0$. Then, the PnL formula of the hedged contract $V_{t}^{H}$
is computed using (\ref{eq:PricingEquationBergomi}) as:

\begin{eqnarray}
\text{PnL}_{V_{t}^{H}} & = & V^{H}(S_{t}+\delta S,\{\xi_{t}^{u}+\delta\xi^{u}\}_{t+\delta t\leq u\leq T_{e}},t+\delta t)-V^{H}(S_{t},\{\xi_{t}^{u}\}_{t\leq u\leq T_{e}},t)\\
 & = & \frac{1}{2}\frac{\partial^{2}V^{H}}{\partial S^{2}}\left(\left(\delta S\right)^{2}-\xi_{t}^{t}S_{t}^{2}\delta t\right)+\int_{t}^{T_{e}}du\frac{\partial^{2}V^{H}}{\partial S\partial\xi^{u}}\left(\delta S\delta\xi^{u}-\rho_{t}^{u}\omega_{t}^{u}\xi_{t}^{u}\sqrt{\xi_{t}^{t}}S_{t}\delta t\right)\nonumber \\
 &  & +\frac{1}{2}\int_{t}^{T_{e}}du\int_{t}^{T_{e}}dv\frac{\partial^{2}V^{H}}{\partial\xi^{u}\partial\xi^{v}}\left(\delta\xi^{u}\delta\xi^{v}-\rho_{t}^{uv}\omega_{t}^{u}\omega_{t}^{v}\xi_{t}^{u}\xi_{t}^{v}\delta t\right)+O(\delta t^{3/2})\label{eq:PnLFormualBergomi}
\end{eqnarray}

(\ref{eq:PnLFormualBergomi}) provides the term-wise break-even condition
for the PnL of $V^{H}$. Therefore, when a non-negligible PnL trend
appears, we can understand easily which term causes that and how we
should modify the model assumptions. Note that all the hedging instrument
price changes can be interpreted as the variations of the state variables
thus we can use the model parameters $\rho_{t}^{u},\rho_{t}^{uv}$
and $\omega_{t}^{u}$ solely for the control of the break-even condition. 
\begin{rem}
In this article, we consider the risk management of a fully hedged
contract $V^{H}$ of an exotic derivative product. In practice, however,
an exotic derivative product is not always fully hedged. For such
case, we regard the partially-hedged contract as the portfolio of
the fully hedged contract $V^{H}$ and the hedging instruments, and
we leave the risk management of the hedging instruments to vanilla
models, which are out of scope of the article.
\end{rem}

\subsection{Setup}

Before proceeding to the discussion about the application of Bergomi
model to interest rate modeling, let us define the basic variables
for the interest rate modeling first. Consider discrete time grids
$T_{i}=\sum_{u=0}^{i-1}\delta_{u},\;T_{0}=0$ with accrual factors
$\{\delta_{u}\}_{u=0,1,...,e-1}$, where $T_{e}$ is the terminal
date of the model. Denote by $P(t,T_{i})=P_{t}^{i}$ the discount
factor at time $t$ with maturity date $T_{i}$. We denote a continuous
bank account process by $B_{t,T}=\exp(\int_{t}^{T}r_{u}du)$ where
$r_{t}$ is a risk-free rate process. Swap rates $S$ and associated
annuity factors $A$ are given as:

\begin{equation}
A_{t}^{i,j}\triangleq\sum_{u=i+1}^{j}\delta_{u-1}P_{t}^{u}
\end{equation}

\begin{equation}
S_{t}^{i,j}\triangleq\frac{P_{t}^{i}-P_{t}^{j}}{A_{t}^{i,j}}
\end{equation}

In the article, we denote $T_{u}$ by $u$ when no confusion can arise,
and we assume that empty sums denote zero and empty products denote
1. 

\subsection{A Variance Swap Contract on an Interest Rate}

Here we introduce an interest rate variance swap contract such that
the variance swap rate will be a martingale under the associated annuity
measure. Owing to the martingale property, we can build the two-factor
model for a swap rate and a forward variance curve under the associated
annuity measure. Also, we are able to compute an European swaption
price using the two-factor model. As we will see later, this simplifies
the relationship between the forward variance curve and the market
European swaption price, and the PnL interpretation in terms of the
market observables becomes clear.

Consider a contract which pays at $T_{l}$ the sum of quadratic variation
of a swap rate $S_{t}^{i,j}$ for the period $[T_{k},T_{l}]$, where
$T_{l}\leq T_{i}$. The quadratic variation is multiplied by $A^{i,j}/P^{l}$
at the end of each observation time grid and rescaled with the factor
$(T_{l}-T_{k})^{-1}$. The payoff at $T_{l}$ is given as:
\begin{equation}
Q_{k,l}^{i,j}=\frac{1}{T_{l}-T_{k}}\sum_{u=k}^{l-1}(S_{u+1}^{i,j}-S_{u}^{i,j})^{2}\frac{A_{u+1}^{i,j}}{P_{u+1}^{l}}
\end{equation}

Also consider a variance swap on a swap rate which pays $Q_{k,l}^{i,j}$
and receives $A_{l}^{i,j}\psi_{t,k,l}^{i,j}$ at $T_{l}$ where $\psi_{t,k,l}^{i,j}$
is fixed at $t$ so that the value of the contract is zero. This means:
\begin{eqnarray}
P_{t}^{l}E_{t}^{T_{l}}[Q_{k,l}^{i,j}-A_{l}^{i,j}\psi_{t,k,l}^{i,j}] & = & \frac{P_{t}^{l}}{T_{l}-T_{k}}\sum_{u=k}^{l-1}E_{t}^{T_{l}}[(S_{u+1}^{i,j}-S_{u}^{i,j})^{2}\frac{A_{u+1}^{i,j}}{P_{u+1}^{l}}]-P_{t}^{l}E_{t}^{T_{l}}[A_{l}^{i,j}\psi_{t,k,l}^{i,j}]\\
 & = & A_{t}^{i,j}\left(\frac{1}{T_{l}-T_{k}}\sum_{u=k}^{l-1}E_{t}^{A^{i,j}}[(S_{u+1}^{i,j}-S_{u}^{i,j})^{2}]-\psi_{t,k,l}^{i,j}\right)=0\label{eq:AnnuityVarSwapParCondition}
\end{eqnarray}

where $E_{t}^{T_{l}}$ and $E_{t}^{A^{i,j}}$denotes $t$-conditional
expectation under $T^{l}$ terminal measure and $A^{i,j}$ annuity
measure respectively.

From (\ref{eq:AnnuityVarSwapParCondition}) we obtain:
\begin{equation}
\psi_{t,k,l}^{i,j}=\frac{1}{T_{l}-T_{k}}\sum_{u=k}^{l-1}E_{t}^{A^{i,j}}[(S_{u+1}^{i,j}-S_{u}^{i,j})^{2}]=\frac{1}{T_{l}-T_{k}}E_{t}^{A^{i,j}}[\left(S_{l}^{i,j}\right)^{2}-\left(S_{k}^{i,j}\right)^{2}]
\end{equation}

This indicates that $\psi_{t,k,l}^{i,j}$ is a martingale under $A^{i,j}$
annuity measure. For $T_{-}<T_{+}$, we get:
\begin{equation}
E_{T_{-}}^{A^{i,j}}[\psi_{T_{+},k,l}^{i,j}]=\frac{1}{T_{l}-T_{k}}E_{T_{-}}^{A^{i,j}}[E_{T_{+}}^{A^{i,j}}[\left(S_{l}^{i,j}\right)^{2}-\left(S_{k}^{i,j}\right)^{2}]]=\frac{1}{T_{l}-T_{k}}E_{T_{-}}^{A^{i,j}}[\left(S_{l}^{i,j}\right)^{2}-\left(S_{k}^{i,j}\right)^{2}]=\psi_{T_{-},k,l}^{i,j}
\end{equation}

\subsection{The Swap Market Bergomi Model}

Next, we consider the dynamics of a swap rate process $S_{t}^{i,j}$.
Because $S_{t}^{i,j}$ is a martingale and hence driftless under $A^{i,j}$
annuity measure, we assume:
\begin{equation}
dS_{t}^{i,j}=a^{i,j}dW_{t}^{(i,j),A^{i,j}}\label{eq:SMBMSDEInit}
\end{equation}

where $a^{i,j}$ is a stochastic process and $W_{t}^{(i,j),A^{i,j}}$
is a Brownian motion under $A^{i,j}$ annuity measure. We define an
infinitesimal variance swap rate $\xi_{t}^{i,j,T}$ as $\lim_{\varepsilon\rightarrow0}\psi_{t,T,T+\varepsilon}^{i,j}$.
This leads to: 

\begin{equation}
\xi_{t}^{i,j,t}=\lim_{\varepsilon\rightarrow0}\psi_{t,t,t+\varepsilon}^{i,j}=\lim_{\varepsilon\rightarrow0}\frac{1}{\varepsilon}E_{t}^{A^{i,j}}[(\int_{t}^{t+\varepsilon}(dS_{u}^{i,j})^{2})]=\lim_{\varepsilon\rightarrow0}\frac{1}{\varepsilon}E_{t}^{A^{i,j}}[\int_{t}^{t+\varepsilon}\left(a_{u}^{i,j}\right)^{2}du]=\left(a_{t}^{i,j}\right)^{2}
\end{equation}

Thus, $a_{t}^{i,j}=\sqrt{\xi_{t}^{i,j,t}}$ and (\ref{eq:SMBMSDEInit})
becomes:
\begin{eqnarray}
dS_{t}^{i,j} & = & \sqrt{\xi_{t}^{i,j,t}}dW_{t}^{(i,j),A^{i,j}}\label{eq:RateSDEAnnuity}
\end{eqnarray}

We assume log-normal dynamics for $\xi_{t}^{i,j,T}$, which is a martingale
under $A^{i,j}$ measure:
\begin{eqnarray}
d\xi_{t}^{i,j,T} & = & \omega^{i,j}e^{-\kappa^{i,j}(T-t)}\xi_{t}^{i,j,T}dZ_{t}^{(i,j),A^{i,j}}
\end{eqnarray}

where $\omega^{i,j}$ and $\kappa^{i,j}$ are static model parameters.

In this article, we assume a variance curve $\{\xi_{t}^{i,j,u}\}_{t<u\leq T_{e}}$
is driven by single Brownian motion $Z_{t}^{(i,j),A^{i,j}}$ for simplicity.
The extension to the multi-factor setting is straightforward. In order
to obtain low-dimensional Markov representation, we introduce the
Ornstein\textendash Uhlenbeck state variable $X_{t}^{i,j}=\int_{0}^{t}e^{-\kappa^{i,j}(t-u)}dZ_{u}^{(i,j),A^{i,j}}$.
With $X_{t}^{i,j}$, the dynamics of $\xi_{t}^{i,j,T}$ is given as:
\begin{eqnarray}
\xi_{t}^{i,j,T} & = & \xi_{0}^{i,j,T}\exp\left[\omega^{i,j}e^{-\kappa^{i,j}(T-t)}X_{t}^{i,j}-\frac{1}{2}\left(\omega^{i,j}\right){}^{2}e^{-2\kappa^{i,j}(T-t)}E^{A^{i,j}}[\left(X_{t}^{i,j}\right)^{2}]\right]\\
dX_{t}^{i,j} & = & -\kappa^{i,j}X_{t}^{i,j}dt+dZ_{t}^{(i,j),A^{i,j}},\;X_{0}^{i,j}=0\label{eq:VolStateSDEAnnuity}
\end{eqnarray}

\section{The Co-terminal Swap Market Bergomi Model}

In the last section, we have built the two-factor model of a swap
rate and a forward variance curve under the associated annuity measure.
In order to evaluate the exotic derivative products which depend on
multiple swap rates like a Bermudan swaption, we need to know the
joint dynamics of the swap rates and the forward variance curves.
In this article, we use the approach of the co-terminal swap market
model \cite{jamshidian1997libor}. 

Using the model, we discuss the second order PnL formula and break-even
levels for a derivative contract with a hedge portfolio. Then, we
analyze the PnL formula using the factor reduction method. Also, we
discuss the computation scheme of the forward variance curves using
the market European swaption prices.

\subsection{Model Dynamics}

The swap market model is classified by the underlying swap rates to
be modeled \cite{jamshidian1997libor,oya2017swap}. In this article,
we consider the co-terminal swap market model for which the yield
curve dynamics is determined by modeling swap rates which share the
common terminal date $T_{e}$ \cite{jamshidian1997libor}. We will
work with $T_{e}$ terminal measure. For the specification of the
co-terminal swap market model, we omit the end index for a swap for
the ease of notation, e.g. $S_{t}^{i,e}=S_{t}^{i}$. We introduce
the state variable vector as $Y\triangleq(S^{1},...,S^{N_{R}},X^{1},...,X^{N_{R}})^{\top}=\{Y^{i}\}_{i=1,...,N_{s}}$
with $N_{R}=e-1,\;N_{s}=2N_{R}$. 

The dynamics of $Y_{t}$ under $T_{e}$ terminal measure are given
by adding the no-arbitrage drifts to (\ref{eq:RateSDEAnnuity}) and
(\ref{eq:VolStateSDEAnnuity}):
\begin{eqnarray}
dY_{t}^{i}=dS_{t}^{i} & = & \sqrt{\xi_{t}^{i,t}}\left(dW_{t}^{(i),T_{e}}+\mu_{t}^{i,T_{e}}dt\right),\;1\leq i\leq N_{R}\label{eq:RateSDETerminal}\\
dY_{t}^{i}=dX_{t}^{i-N_{R}} & = & -\kappa^{i-N_{R}}X_{t}^{i-N_{R}}dt+\left(dW_{t}^{(i),T_{e}}+\mu_{t}^{i,T_{e}}dt\right),\;N_{R}<i\leq N_{S}\label{eq:VolSDETerminal}
\end{eqnarray}

where $\{W_{t}^{(i),T_{e}}\}_{i=1,...,N_{s}}$ are the Brownian motions
under $T_{e}$ terminal measure defined as:
\begin{eqnarray}
W_{t}^{(i),T_{e}}+\int_{0}^{t}\mu_{u}^{i,T_{e}}du & =\begin{cases}
\begin{array}{ccc}
W_{t}^{(i),A^{i}}, &  & 1\leq i\leq N_{R}\\
Z_{t}^{(i-N_{R}),A^{i-N_{R}}}, &  & N_{R}<i\leq N_{S}
\end{array}\end{cases}
\end{eqnarray}
We denote the correlation functions as:
\begin{eqnarray}
\langle dW^{(i),T_{e}},dW^{(j),T_{e}}\rangle_{t} & = & \rho_{t}^{Y,ij}dt
\end{eqnarray}

\begin{prop}
No-arbitrage drifts in (\ref{eq:RateSDETerminal})-(\ref{eq:VolSDETerminal})
are given as:
\begin{eqnarray}
\mu_{t}^{i,T_{e}} & = & -\sum_{u=a(i)+1}^{e-1}\frac{s_{t}^{a(i)u}}{s_{t}^{a(i)}}\frac{\delta_{u-1}\sqrt{\xi_{t}^{u,t}}\rho_{t}^{Y,iu}}{1+\delta_{u-1}S_{t}^{u}},\;1\leq i\leq N_{S}
\end{eqnarray}
with:
\begin{eqnarray}
a(i) & = & \begin{cases}
\begin{array}{ccc}
i, &  & 1\leq i\leq N_{R}\\
i-N_{R}, &  & N_{R}<i\leq N_{S}
\end{array}\end{cases}\\
s_{t}^{ij} & \triangleq & \sum_{u=j}^{e-1}\delta_{u}\prod_{v=i+1}^{u}(1+\delta_{v-1}S_{t}^{v})\\
s_{t}^{i} & \triangleq & s_{t}^{ii}
\end{eqnarray}
\end{prop}
The proof is given in the Appendix.

\subsection{The PnL Formula}

Here we analyze the PnL formula with the co-terminal SMBM and discuss
how we could adjust the model parameters using the results of the
PnL analysis. Assume that we hold a derivative contract and the value
of the contract is denoted by $V(Y_{t},t)$. The pricing equation
of $V(Y_{t},t)$ is given as:
\begin{equation}
\frac{\partial V}{\partial t}-r_{t}V+\sum_{u=1}^{N_{s}}\frac{\partial V}{\partial Y^{u}}\mu_{t}^{u,Q}+\frac{1}{2}\sum_{u,v=1}^{N_{s}}\frac{\partial^{2}V}{\partial Y^{u}\partial Y^{v}}\rho_{t}^{Y,uv}\sigma_{t}^{Y,u}\sigma_{t}^{Y,v}=0\label{eq:PricingEqCoterminalSMBM}
\end{equation}
where $\mu_{t}^{u,Q}$ is the no-arbitrage drift of $Y^{u}$ under
the risk-neutral measure. We don't provide the explicit form of $\mu_{t}^{u,Q}$
here because it will be irrelevant for the discussion of a hedged
contract. $\sigma_{t}^{Y,i}$ is given as:
\begin{eqnarray}
\sigma_{t}^{Y,i} & =\begin{cases}
\begin{array}{ccc}
\sqrt{\xi_{t}^{i,t}}, &  & 1\leq i\leq N_{R}\\
1, &  & N_{R}<i\leq N_{S}
\end{array}\end{cases}
\end{eqnarray}

Next, we consider a hedged contract $V^{H}$ for the derivative contract:
\begin{equation}
V^{H}=V+\sum_{u=1}^{N_{s}}w_{u}H^{u}+w_{N_{S}+1}B\label{eq:HedgedPortfolio}
\end{equation}
where $\{w_{u}\}_{1\leq u\leq N_{S}+1}$ are the hedge weights. The
prices of hedging instruments are given as:
\begin{equation}
H_{t}^{i}=\begin{cases}
\begin{array}{ccc}
A_{t}^{i}(K^{i}-S_{t}^{i}), & \text{} & 1\leq i\leq N_{R}\\
A_{t}^{i}E_{t}^{A^{i}}[(S_{i}^{i}-K^{i})^{+}], & \text{} & N_{R}<i\leq N_{s}
\end{array}\end{cases}\label{eq:HedgeInstruments}
\end{equation}

Namely, we hedge the derivative contract with interest rate swaps,
payers European swaptions of strike $K^{i}$ and a bank account $B$.
We build the hedged contract so as to satisfy the below equations:
\begin{eqnarray}
\frac{\partial V^{H}}{\partial Y^{i}} & = & \frac{\partial V}{\partial Y^{i}}+\sum_{u=1}^{N_{s}}w_{u}\frac{\partial H^{u}}{\partial Y^{i}}=0,\;1\leq i\leq N_{s}\label{eq:HedgeConditionDelta}\\
V^{H} & = & 0\label{eq:HedgeConditionValue}
\end{eqnarray}

The hedge conditions (\ref{eq:HedgeConditionDelta}) and (\ref{eq:HedgeConditionValue})
are given as a linear system and solved using standard linear algebra.

We consider the second order PnL formula for the hedged contract $V^{H}$.
Using the Taylor expansion of order 2 in $\delta Y$, we obtain:
\begin{eqnarray}
\text{PnL}_{V^{H}} & \triangleq & V^{H}(Y_{t}+\delta Y,t+\delta t)-V^{H}(Y_{t},t)\\
 & = & \frac{\partial V^{H}}{\partial t}\delta t+\frac{1}{2}\sum_{u,v=1}^{N_{s}}\frac{\partial^{2}V^{H}}{\partial Y^{u}\partial Y^{v}}\delta Y^{u}\delta Y^{v}+O(\delta t^{3/2})\\
 & = & \frac{1}{2}\sum_{u,v=1}^{N_{s}}\frac{\partial^{2}V^{H}}{\partial Y^{u}\partial Y^{v}}\left(\delta Y^{u}\delta Y^{v}-\rho_{t}^{Y,uv}\sigma_{t}^{Y,u}\sigma_{t}^{Y,v}\delta t\right)+O(\delta t^{3/2})\label{eq:HedgedSecondOrderPnL}
\end{eqnarray}

(\ref{eq:HedgedSecondOrderPnL}) indicates the second order PnL is
given as the sum of the differences of the realized quadratic cross
variations $\delta Y^{u}\delta Y^{v}$ and the break-even level $\rho_{t}^{Y,uv}\sigma_{t}^{Y,u}\sigma_{t}^{Y,v}\delta t$
multiplied by the gamma term $\frac{\partial^{2}V^{H}}{\partial Y^{u}\partial Y^{v}}$
just as (\ref{eq:PnLFormualBergomi}). Therefore, we can carry out
the term-wise analysis in the same manner as the Bergomi model. The
calibration procedure is not requried for the co-terminal SMBM and
we can use the model parameters solely for the control of the break-even
levels. Let us summarize the model parameters as follows:
\begin{itemize}
\item $\xi_{0}^{i,T}$ defines the shape of the initial forward variance
curve, that affects $\sqrt{\xi_{t}^{i,t}}$ which appears in the break-even
level expression as $\sigma_{t}^{Y,i}$ for $1\leq i\leq N_{R}$.
As we will see later, we will set a constraint for $\xi_{0}^{i,T}$
such that the model will reproduce a market European swaption price
at $t=0$. Even with that constraint, we still have freedom in $\xi_{0}^{i,T}$
to control the break-even level.
\item $\omega^{i}$ controls the log-normal volatility of an infinitesimal
forward variance swap rates $\xi_{t}^{i,T}$. In terms of the state
variable, $\omega^{i}$ controls the scale of $X^{i}$. Because $\sigma_{t}^{Y,i}=1$
always holds for $N_{R}<i\leq N_{s}$, the control of the scale of
$X^{i}$ is equivalent to the control of the break-even levels for
$\delta X^{i}$.
\item $\rho_{t}^{Y,ij}$ controls the correlations between state variables.
\end{itemize}

\subsection{Factor Analysis of the PnL}

Because an interest rate model often contains a large number of state
variables, practitioners frequently use the model driven by a limited
number of Brownian motions so as to reduce the complexity. The approach
is also useful with the co-terminal SMBM for the intuitive understanding
of the economy of the product. We will consider the co-terminal SMBM
for which swap rates and forward variance curves are driven by three
Brownian motions:

\begin{equation}
dW_{t}^{(i),T_{e}}=\begin{cases}
\begin{array}{ccc}
c^{(i),1}dW_{t}^{C,1}+c^{(i),2}dW_{t}^{C,2}, &  & 1\leq i\leq N_{R}\\
dW_{t}^{C,3}, &  & N_{R}<i\leq N_{S}
\end{array}\end{cases}\label{eq:FactorReduction}
\end{equation}

where we parameterize $\{c^{(i),u}\}_{u=1,2}$ as $c^{(i),1}=\cos\alpha^{(i)}$,
$c^{(i),2}=\sin\alpha^{(i)}$. We assume the correlations are given
as $\langle dW^{C,i},dW^{C,j}\rangle_{t}=\rho_{t}^{C,ij}dt$ for $i,j=1,2,3$
with $\rho_{t}^{C,12}=0$. 

We consider the PnL formula associated with the above Brownian motion
changes. Define the value function of a derivative contract parameterized
with the reduced factors as $V^{h}(h,t)\triangleq V(Y_{t}+\sum_{u=1}^{3}h^{u}m^{u},t)$
with $h=\{h^{u}\in\mathbb{R}\}_{u=1,2,3}$ and $m=\{m^{u}\in\mathbb{R}^{N_{S}}\}_{u=1,2,3}$.
$m$ is given as:
\begin{eqnarray}
m^{1} & = & (\sqrt{\xi_{t}^{1,t}}c^{(1),1},...,\sqrt{\xi_{t}^{N_{R},t}}c^{(N_{R}),1},\underbrace{0,...,0}_{N_{R}})^{\top}\\
m^{2} & = & (\sqrt{\xi_{t}^{1,t}}c^{(1),2},...,\sqrt{\xi_{t}^{N_{R},t}}c^{(N_{R}),2},\underbrace{0,...,0}_{N_{R}})^{\top}\\
m^{3} & = & (\underbrace{0,...,0}_{N_{R}},\underbrace{1,...,1}_{N_{R}})^{\top}
\end{eqnarray}
With $V^{h}$, we can write the theta term for the hedged contract
in (\ref{eq:HedgedSecondOrderPnL}) as:
\begin{eqnarray}
\frac{1}{2}\sum_{u,v=1}^{N_{s}}\rho_{t}^{Y,uv}\sigma_{t}^{Y,u}\sigma_{t}^{Y,v}\frac{\partial^{2}V^{H}}{\partial Y^{u}\partial Y^{v}}\delta t & = & \frac{1}{2}\sum_{u,v=1}^{N_{s}}\sum_{i,j=1}^{3}\rho_{t}^{C,ij}\left(m^{i}\right)_{u}\left(m^{j}\right)_{v}\frac{\partial^{2}V^{H}}{\partial Y^{u}\partial Y^{v}}\delta t\\
 & = & \frac{1}{2}\sum_{i,j=1}^{3}\rho^{C,ij}\frac{\partial^{2}V^{H,h}}{\partial h^{i}\partial h^{j}}\delta t\label{eq:ComponentCarry}
\end{eqnarray}

where $V^{H,h}\triangleq V^{H}(Y_{t}+\sum_{u=1}^{3}h^{u}m^{u},t)$.
Also, the gamma term in (\ref{eq:HedgedSecondOrderPnL}) becomes:
\begin{equation}
\frac{1}{2}\sum_{u,v=1}^{N_{s}}\frac{\partial^{2}V^{H}}{\partial Y^{u}\partial Y^{v}}\delta Y^{u}\delta Y^{v}=\frac{1}{2}\sum_{i,j=1}^{3}\frac{\partial^{2}V^{H,h}}{\partial h^{i}\partial h^{j}}\delta h^{i}\delta h^{j}+\frac{1}{2}\sum_{u,v=1}^{N_{s}}\frac{\partial^{2}V^{H}}{\partial Y^{u}\partial Y^{v}}\left(\delta Y^{u}\delta Y^{v}-\sum_{i,j=1}^{3}\left(m^{i}\right)_{u}\left(m^{j}\right)_{v}\delta h^{i}\delta h^{j}\right)\label{eq:ComponentGamma}
\end{equation}

where $\delta h=\{\delta h^{u}\in\mathbb{R}\}_{u=1,2,3}$ is defined
so that the $L_{2}$-norm of $\delta Y-\sum_{u=1}^{3}\delta h^{u}m^{u}$
is minimized:
\begin{equation}
\delta h=\text{argmin}_{p}||\delta Y-\sum_{u=1}^{3}p^{u}m^{u}||_{2}\label{eq:dh_compute}
\end{equation}
With (\ref{eq:ComponentCarry}) and (\ref{eq:ComponentGamma}), we
obtain the second order PnL formula with the reduced factors:
\begin{equation}
\frac{1}{2}\sum_{i,j=1}^{3}\frac{\partial^{2}V^{H,h}}{\partial h^{i}\partial h^{j}}\left(\delta h^{i}\delta h^{j}-\rho_{t}^{C,ij}\delta t\right)+\frac{1}{2}\sum_{u,v=1}^{N_{s}}\frac{\partial^{2}V^{H}}{\partial Y^{u}\partial Y^{v}}\left(\delta Y^{u}\delta Y^{v}-\sum_{i,j=1}^{3}\left(m^{i}\right)_{u}\left(m^{j}\right)_{v}\delta h^{i}\delta h^{j}\right)\label{eq:PnLComponent}
\end{equation}
If the model with the reduced factors is able to reproduce the realized
dynamics of the state variables well, the contribution from the second
term of (\ref{eq:PnLComponent}) will be negligible in comparison
to the first term. If it is not the case, we should review the assumption
of the factor reduction. Here we assume that the reduced factors can
reproduce the realized dynamics well and the PnL can be explained
accurately enough with the first term of (\ref{eq:PnLComponent}).
Then, to understand the PnL of the derivative contract, we should
check which term of $\frac{\partial^{2}V^{H,h}}{\partial h^{i}\partial h^{j}}$
is large, and the break-even level $\rho_{t}^{C,ij}\delta t$ is reasonable
in comparison to the realized quadratic cross variation $\delta h^{i}\delta h^{j}$,
just in the same manner as standard PnL analysis. The benefit from
the factor reduction is that we need to have a look at much less number
of the gamma term and the break-even levels, which is 6 in our setting.

\subsection{Forward Variance Curve Computation with Swaption Market Prices}

Next, we discuss how we compute the forward variance curve $\xi_{t}^{i,T}$
using market prices of European swaptions. We use the different computation
procedures for $t=0$ and $t>0$. At $t=0$, the initial forward variance
curve $\xi_{0}^{i,T}$ is computed in order for the model to reproduce
the market prices of European swaptions assuming $X_{t}^{i}=0$. For
$t>0$, European swaption market price changes are reflected in the
state variable $X_{t}^{i}$. In other words, the model parameter calibration
to market prices of European swaptions are performed only at $t=0$,
and after that the market price changes are interpreted as the model
state variables changes. As a result, the PnL is not affected by the
diffusion parameter changes and the interpretation of the PnL will
become simple.

The initial forward variance curve is parameterized as below:
\begin{equation}
\xi_{0}^{i,T}=\left(\sigma_{0}^{i}\right)^{2}e^{\theta^{i}T},\;1\leq i\leq N_{R}
\end{equation}

$\theta^{i}$ is the model parameter for the control of the initial
forward variance curve shape $\xi_{0}^{i,T}$. $\sigma_{0}^{i}$ is
calibrated to an European swaption market price at $t=0$ and then
fixed for $t>0$.

Consider the computation of $\xi_{t}^{i,T}$ at $t=T_{s}$. We will
obtain $\xi_{s}^{i,T}$ so that the model reproduce the market price
of an European swaption on $S^{i}$ with strike $K$ using the efficient
numerical computation scheme presented in \cite{bergomi2015stochastic}.
Here we use the two-factor model for a swap rate $S^{i}$ and a state
variable $X^{i}$ under the associated annuity measure for the computation
of the model swaption price. Denote by $T_{i}$ expiry time of a swaption
on $S^{i}$. Note $S^{i}$ become driftless under $A^{i}$ annuity
measure:
\begin{eqnarray}
dS_{t}^{i} & = & \sqrt{\xi_{t}^{i,t}}dW_{t}^{(i),A^{i}}=\sqrt{\xi_{t}^{i,t}}\left(\rho_{t}^{Y,i(i+N_{R})}dZ_{t}^{(i),A^{i}}+\sqrt{1-\left(\rho_{t}^{Y,i(i+N_{R})}\right)^{2}}dZ_{t}^{(i),\perp,A^{i}}\right)\label{eq:SwapRateSDECalib}\\
dX_{t}^{i} & = & -\kappa^{i}X_{t}^{i}dt+dZ_{t}^{(i),A^{i}}
\end{eqnarray}

where $Z_{t}^{(i),\perp,A^{i}}$ is a Brownian motion which is independent
from $Z_{t}^{(i),A^{i}}$. 

Next, let us look at a payers swaption price on a swap rate $S^{i}$,
$V_{s}^{i}=A_{s}^{i}E_{s}^{A^{i}}[(S_{i}^{i}-K)^{+}]$. Because $Z_{t}^{(i),\perp,A^{i}}$
and $Z_{t}^{(i),A^{i}}$ are independent, the expectation can be written
using the tower rule as:
\begin{equation}
E_{s}^{A^{i}}[(S_{i}^{i}-K)^{+}]=E_{s}^{A^{i}}[E_{s}^{A^{i}}[(S_{i}^{i}-K)^{+}|\{Z_{u}^{(i),A^{i}}\}_{T_{s}\leq u\leq T_{i}}]]\label{eq:TowerRulePayersExpectation}
\end{equation}

From (\ref{eq:SwapRateSDECalib}) we can see $S_{i}^{i}$ conditioned
on a path for $Z_{t}^{(i),A^{i}}$ follows normal distribution with
the moments:
\begin{eqnarray}
E_{s}^{A^{i}}[S_{i}^{i}|\{Z_{u}^{(i),A^{i}}\}_{T_{s}\leq u\leq T_{i}}] & = & S_{s}^{i}+\int_{T_{s}}^{T_{i}}\sqrt{\xi_{u}^{i,u}}\rho_{u}^{Y,i(i+N_{R})}dZ_{u}^{(i),A^{i}}\triangleq\bar{S}^{i}\\
E_{s}^{A^{i}}[(S_{i}^{i}-\bar{S}^{i})^{2}|\{Z_{u}^{(i),A^{i}}\}_{T_{s}\leq u\leq T_{i}}] & = & \int_{T_{s}}^{T_{i}}\xi_{u}^{i,u}\left(1-\left(\rho_{u}^{Y,i(i+N_{R})}\right)^{2}\right)du\triangleq\left(\hat{\sigma}^{i}\right)^{2}(T_{i}-T_{s})
\end{eqnarray}

Thus the inner expectation of (\ref{eq:TowerRulePayersExpectation})
can be calculated analytically using the normal Black-Scholes formula:
\begin{eqnarray}
E_{s}^{A^{i}}[(S_{i}^{i}-K)^{+}|\{Z_{u}^{(i),A^{i}}\}_{T_{s}\leq u\leq T_{i}}] & = & NBS(\bar{S}^{i},\hat{\sigma}^{i},K,T_{i}-T_{s})\\
NBS(S,\sigma,K,t) & \triangleq & (S-K)N(\frac{S-K}{\sigma\sqrt{t}})+\sigma\sqrt{t}N^{'}(\frac{S-K}{\sigma\sqrt{t}})
\end{eqnarray}

where $N(x)$ is the standard cumulative normal distribution function
and $N^{'}(x)=dN/dx$. Then, the swaption price is obtained as:
\begin{equation}
A_{s}^{i}E^{A^{i}}[(S_{i}^{i}-K)^{+}]=A_{s}^{i}E_{s}^{A^{i}}[NBS(\bar{S}^{i},\hat{\sigma}^{i},K,T_{i}-T_{s})]\label{eq:CalibCondWithPrice}
\end{equation}

We assume that the market swaption price for an European swaption
on $S^{i}$ as of $T_{s}$ is given as a normal implied volatility
$\sigma_{s}^{I,i}$. We require the model price of the swaption (\ref{eq:CalibCondWithPrice})
to be equal to the market price:
\begin{equation}
E_{s}^{A^{i}}[NBS(\bar{S}^{i},\hat{\sigma}^{i},K,T_{i}-T_{s})]=NBS(S_{s}^{i},\sigma_{s}^{I,i},K,T_{i}-T_{s})\label{eq:CalibrationCondition}
\end{equation}

(\ref{eq:CalibrationCondition}) is solved using a one-dimensional
root-finding algorithm, in terms of $\sigma_{0}^{i}$ for $T_{s}=0$
and $X_{s}^{i}$ for $T_{s}>0$. 
\begin{rem}
Owing to the martingale property of $\xi_{t}^{i,T}$ under $A^{i}$
annuity measure, the expectation $E_{t}^{A^{i}}[(S_{i}^{i}-K)^{+}]$
depends on the dynamics of $S_{t}^{i}$ and $\xi_{t}^{i,T}$ but not
on $\{S_{t}^{u}\}_{u\neq i}$ or $\{\xi_{t}^{u,T}\}_{u\neq i}$. In
consequence, the state variable is related to market observables in
a simple manner, namely $X_{t}^{i}$ (and equivalently forward variance
curve $\xi_{t}^{i,T}$) relies on $S_{t}^{i}$ and $\sigma_{t}^{I,i}$
but not on $\{S_{t}^{u}\}_{u\neq i}$ or $\{\sigma_{t}^{I,u}\}_{u\neq i}$.
\end{rem}

\section{A Numerical Experiment with The Co-terminal SMBM}

Bermudan swaptions have been traded for a long time and one of the
most popular exotic interest rate products. However, the position
management of the product is still a challenging task. In this section,
we perform a numerical experiment for a Canary swaption, which is
a Bermudan swaption with only two exercise dates, and illustrate how
the PnL analysis can be performed in a systematic manner using the
co-terminal SMBM.

\subsection{Canary Swaption Pricing}

We assume the two interest rate swaps underlying a Canary swaption
have the common terminal date $T_{e}$. The price of a receiver Canary
swaption $V^{C}$ is given as below:
\begin{eqnarray}
V^{C} & = & P_{0}^{e}E_{0}^{T_{e}}[\frac{\max(U_{i_{1}},O_{i_{1}})}{P_{i_{1}}^{e}}]\\
U_{i_{1}} & = & A_{i_{1}}^{i_{1}}(K-S_{i_{1}}^{i_{1}})\\
O_{i_{1}} & = & A_{i_{1}}^{i_{2}}E_{i_{1}}^{A^{i_{2}}}[(K-S_{i_{2}}^{i_{2}})^{+}]
\end{eqnarray}

where $\{i_{u}\}_{u=1,2},\;i_{1}<i_{2}$ is the time indices for the
start time of underlying swap rates and $K$ is the strike. Because
$O_{i_{1}}$ is given as the conditional expectation at $T_{i_{1}}$,
the exact simulation requires a nested Monte Carlo method that takes
a huge amount of valuation time. Therefore, in practice, an approximation
method is often applied for the valuation of Bermudan swaptions like
the least square Monte Carlo method, which we use for this analysis.
Besides, in order to see the model behaviour with less numerical errors,
we use a semi-nested Monte Carlo method described as below:
\begin{enumerate}
\item We sample $(S_{i_{1}}^{i_{2}},X_{i_{1}}^{i_{2}})$ for reasonably
wide range of values. For each sample of $(S_{i_{1}}^{i_{2}},X_{i_{1}}^{i_{2}})$,
we perform Monte Carlo simulation to compute $E_{i_{1}}^{A^{i_{2}}}[(K-S_{i_{2}}^{i_{2}})^{+}]$
and calculate the implied normal volatility $\sigma_{i_{1}}^{I,i_{2}}$.
Then we build a spline function $\sigma_{i_{1}}^{I,i_{2}}=g(S_{i_{1}}^{i_{2}},X_{i_{1}}^{i_{2}})$
using the samples. 
\item In Monte Carlo simulation for pricing, we diffuse the model up to
$T_{i_{1}}$, and we obtain $\sigma_{i_{1}}^{I,i_{2}}$ as $g(S_{i_{1}}^{i_{2}},X_{i_{1}}^{i_{2}})$.
Then, we evaluate $E_{i_{1}}^{A^{i_{2}}}[(K-S_{i_{2}}^{i_{2}})^{+}]$
using normal Black-Scholes formula with $S_{i_{1}}^{i_{2}}$, $X_{i_{1}}^{i_{2}}$
and $\sigma_{i_{1}}^{I,i_{2}}=g(S_{i_{1}}^{i_{2}},X_{i_{1}}^{i_{2}})$. 
\end{enumerate}
We use the reduced factor model (\ref{eq:FactorReduction}) with the
common parameters among the swap rates, $\theta^{i}=\theta$, $\omega^{i}=\omega$
and $\kappa^{i}=\kappa$. The correlation functions are parameterized
as following:
\begin{eqnarray}
\rho^{C,23} & = & 0\\
\rho^{RV} & \triangleq & \rho^{C,13}\\
\rho^{RR} & \triangleq & \cos\alpha^{(i_{1})}=c^{(i_{1}),1},\;0\leq\alpha^{(i_{1})}\leq\pi\label{eq:RRCorrDef}\\
\alpha^{(e-1)} & = & -\alpha^{(i_{1})}\\
\alpha^{(j)} & = & \frac{T_{j}-T_{i_{1}}}{T_{e-1}-T_{i_{1}}}\alpha^{(e-1)}+\frac{T_{e-1}-T_{j}}{T_{e-1}-T_{i_{1}}}\alpha^{(i_{1})},\;i_{1}<j<e-1
\end{eqnarray}

Monte Carlo simulation are performed with the quasi-Monte Carlo method
with the number of paths of $2^{17}$. 

\subsection{Numerical Exercise of PnL Analysis}

We examined the PnL for a hedged Canary swaption $V^{CH,h}$ where
$V^{CH}$ is the price of the hedged contract (\ref{eq:HedgedPortfolio}),(\ref{eq:HedgeInstruments})
for $V^{C}$ and $V^{CH,h}(h,t)\triangleq V^{CH}(Y_{t}+\sum_{u=1}^{3}h^{u}m^{u},t)$
is parameterized with the reduced factors using $h$ and $m$ as in
the section 3.3. Note the hedged contract $V^{CH}$ is constructed
at $t=0$ and the hedge weights are unchanged for $[0,\delta t]$.
We assumed that the time difference was $\delta t=0.01$ and the state
variable differences were fully expressed with the reduced factors,
namely $\delta Y=\sum_{u=1}^{3}\delta h^{u}m^{u}$ with $\delta h=\{0.12,0.08,0.08\}$.
We analysed the realized PnL $V^{CH,h}(\delta h,\delta t)-V^{CH,h}(0,0)$
using the below formula:
\begin{equation}
V^{CH,h}(\delta h,\delta t)-V^{CH,h}(0,0)\sim\sum_{i,j=1}^{3}\frac{1}{2}\frac{\partial^{2}V^{CH,h}}{\partial h^{i}\partial h^{j}}\left(\delta h^{i}\delta h^{j}-\rho_{t}^{C,ij}\delta t\right)\label{eq:PnLAnalysisFormula}
\end{equation}

The table 2 shows the analysis results with the semi-nested and the
least square Monte Carlo method. Firstly, we can confirm that the
approximation (\ref{eq:PnLAnalysisFormula}) worked fine for the semi-nested
Monte Carlo method; out of the total realized PnL of -0.48 basis point,
-0.45 basis point was explained by the right hand side of (\ref{eq:PnLAnalysisFormula}).
Secondly, we can clearly understand why the non-zero realized PnL
$V^{CH,h}(\delta h,\delta t)-V^{CH,h}(0,0)$ appeared; the covariance
dynamics of $\delta h^{i}$ and $\delta h^{j}$ for $(i,j)=(1,1),(1,2),(2,1)$
disagreed between the model assumptions $\rho_{t}^{C,ij}\delta t$
and the realized numbers $\delta h^{i}\delta h^{j}$ then these PnL
components contributed most to the non-zero realized PnL. If this
PnL trend persists for a while, we may better to change the model
parameters for such covariance pairs. For example, to make the PnL
of $(i,j)=(1,1)$ component close to flat, we need to assume larger
values for $\{\xi_{t}^{u,t}\}_{i_{1}\leq u\leq e-1}$, which means
we are required to decrease $\theta$ and then re-compute $\{\sigma_{0}^{u}\}_{i_{1}\leq u\leq e-1}$.
As for the result with the least square Monte Carlo method, the overall
behaviour was similar to the semi-nested Monte Carlo method, but we
saw a bit larger unexplained PnL. This would be due to the numerical
error arising from the least square approximation of the exercise
boundary. A numerical scheme of the least square Monte Carlo method
with higher accuracy for the SMBM is the subject of the future study.

\begin{table}
\begin{tabular}{|c||c|c|c|c|c|c|c|c|c|}
\hline 
$T_{i}$ & 1 & 2 & 3 & 4 & 5 & 6 & 7 & 8 & 9\tabularnewline
\hline 
\hline 
$S_{0}^{i}$ & 2.53\% & 2.57\% & 2.57\% & 2.57\% & 2.58\% & 2.59\% & 2.60\% & 2.60\% & 2.62\%\tabularnewline
\hline 
$\sigma_{0}^{I,i}$ & 0.658\% & 0.698\% & 0.718\% & 0.729\% & 0.739\% & 0.740\% & 0.739\% & 0.736\% & 0.724\%\tabularnewline
\hline 
\end{tabular}

\caption{The forward swap rates and implied normal volatilities}
\end{table}

\begin{table}
The semi-nested Monte Carlo method:

\begin{tabular}{|c|c|c|c|c|}
\hline 
($i,j$) & $\frac{\partial^{2}V^{CH,h}}{\partial h^{i}\partial h^{j}}$ & $\frac{1}{2}\frac{\partial^{2}V^{CH,h}}{\partial h^{i}\partial h^{j}}\delta h^{i}\delta h^{j}$ & $-\frac{1}{2}\frac{\partial^{2}V^{CH,h}}{\partial h^{i}\partial h^{j}}\rho_{t}^{C,ij}\delta t$ & $\frac{1}{2}\frac{\partial^{2}V^{CH,h}}{\partial h^{i}\partial h^{j}}\left(\delta h^{i}\delta h^{j}-\rho_{t}^{C,ij}\delta t\right)$\tabularnewline
\hline 
\hline 
(1,1) & -0.765\% & -0.55e-4 & +0.38e-4 & -0.17e-4\tabularnewline
\hline 
(2,2) & +0.124\% & +0.04e-4 & -0.06e-4 & -0.02e-4\tabularnewline
\hline 
(3,3) & -0.011\% & -0.00e-4 & +0.01e-4 & +0.00e-4\tabularnewline
\hline 
(1,2),(2,1) & -0.199\% & -0.10e-4 & +0.00e-4 & -0.10e-4\tabularnewline
\hline 
(1,3),(3,1) & -0.084\% & -0.04e-4 & +0.01e-4 & -0.03e-4\tabularnewline
\hline 
(2,3),(3,2) & -0.017\% & -0.01e-4 & +0.00e-4 & -0.01e-4\tabularnewline
\hline 
\hline 
\multicolumn{2}{|c|}{$\sum_{i,j=1}^{3}$} & -0.80e-4 & +0.34e-4 & -0.45e-4\tabularnewline
\hline 
\multicolumn{1}{c}{} & \multicolumn{1}{c}{} & \multicolumn{1}{c}{} & \multicolumn{1}{c}{} & \multicolumn{1}{c}{}\tabularnewline
\cline{5-5} 
\multicolumn{1}{c}{} & \multicolumn{1}{c}{} & \multicolumn{1}{c}{} &  & $V^{CH,h}(\delta h,\delta t)-V^{CH,h}(0,0)$\tabularnewline
\cline{4-5} 
\multicolumn{2}{c}{} &  & Realized PnL & -0.48e-4\tabularnewline
\cline{4-5} 
\multicolumn{1}{c}{} & \multicolumn{1}{c}{} & \multicolumn{1}{c}{} & \multicolumn{1}{c}{} & \multicolumn{1}{c}{}\tabularnewline
\end{tabular}

The least square Monte Carlo method:

\begin{tabular}{|c|c|c|c|c|}
\hline 
($i,j$) & $\frac{\partial^{2}V^{CH,h}}{\partial h^{i}\partial h^{j}}$ & $\frac{1}{2}\frac{\partial^{2}V^{CH,h}}{\partial h^{i}\partial h^{j}}\delta h^{i}\delta h^{j}$ & $-\frac{1}{2}\frac{\partial^{2}V^{CH,h}}{\partial h^{i}\partial h^{j}}\rho_{t}^{C,ij}\delta t$ & $\frac{1}{2}\frac{\partial^{2}V^{CH,h}}{\partial h^{i}\partial h^{j}}\left(\delta h^{i}\delta h^{j}-\rho_{t}^{C,ij}\delta t\right)$\tabularnewline
\hline 
\hline 
(1,1) & -0.801\% & -0.58e-4 & +0.40e-4 & -0.18e-4\tabularnewline
\hline 
(2,2) & +0.133\% & +0.04e-4 & -0.07e-4 & -0.02e-4\tabularnewline
\hline 
(3,3) & -0.011\% & -0.00e-4 & +0.01e-4 & +0.00e-4\tabularnewline
\hline 
(1,2),(2,1) & -0.211\% & -0.10e-4 & +0.00e-4 & -0.10e-4\tabularnewline
\hline 
(1,3),(3,1) & -0.088\% & -0.04e-4 & +0.01e-4 & -0.03e-4\tabularnewline
\hline 
(2,3),(3,2) & -0.018\% & -0.01e-4 & +0.00e-4 & -0.01e-4\tabularnewline
\hline 
\hline 
\multicolumn{2}{|c|}{$\sum_{i,j=1}^{3}$} & -0.84e-4 & +0.36e-4 & -0.48e-4\tabularnewline
\hline 
\multicolumn{1}{c}{} & \multicolumn{1}{c}{} & \multicolumn{1}{c}{} & \multicolumn{1}{c}{} & \multicolumn{1}{c}{}\tabularnewline
\cline{5-5} 
\multicolumn{1}{c}{} & \multicolumn{1}{c}{} & \multicolumn{1}{c}{} &  & $V^{CH,h}(\delta h,\delta t)-V^{CH,h}(0,0)$\tabularnewline
\cline{4-5} 
\multicolumn{2}{c}{} &  & Realized PnL & -0.58e-4\tabularnewline
\cline{4-5} 
\end{tabular}

\caption{PnL analysis result for a Canary swaption with $T_{i_{1}}=1$, $T_{i_{2}}=4$,
$T_{e}=10$, $K=3.00\%$. The strike for all the hedging interest
rate swaps and swaptions are also $3.00\%$. The model parameters
are set to $\theta=0.0$, $\omega=0.3$, $\kappa=0.1$, $\rho^{RR}=0.9$,
$\rho^{RV}=0.2$, and $P_{0}^{1}=0.975$. The accrual factors are
assumed to be flat, $\delta_{u}=1.0$ for $i_{1}\leq u\leq e-1$.
The forward rates and implied normal volatilities are given in the
Table 1, with which $\{\sigma_{0}^{u}\}_{i_{1}\leq u\leq e-1}$ is
computed using (\ref{eq:CalibrationCondition}) at $t=0$. }
\end{table}

\section{Conclusion}

In this article, we applied the forward variance modeling approach
to the co-terminal swap market model. The model has an advantage that
the diffusion parameter calibration is not required to take the market
price changes of the hedging instruments into account. As a result,
the PnL formula of a hedged contract becomes quite simple. We numerically
illustrated how the PnL analysis was performed with the co-terminal
SMBM for a Canary swaption and confirmed that we could clearly understand
the PnL and easily determine which model parameter to be changed to
cope with non-flat PnL.

\section{Disclaimer}

The author would like to sincerely thank Kei Minakuchi and Paul McCloud
for their support that greatly contributed to the improvement of the
article. The opinions expressed in this article are the author's own
and do not reflect the view of Nomura Securities Co., Ltd. All errors
are the author's responsibility.

\section*{Appendix}

\subsection*{Proof of Proposition 1}
\begin{proof}
First, we prove the below equation for $s_{t}^{i}$ by induction:
\begin{equation}
A_{t}^{i}=P_{t}^{e}s_{t}^{i},\;1\leq i\leq e-1\label{eq:Proposition1_Appendix}
\end{equation}
We compute $P_{t}^{e}s_{t}^{e-1}$ as:
\begin{eqnarray}
P_{t}^{e}s_{t}^{e-1} & = & P_{t}^{e}\delta_{e-1}=A_{t}^{e-1}\label{eq:InductionFirst}
\end{eqnarray}
(\ref{eq:InductionFirst}) indicates (\ref{eq:Proposition1_Appendix})
holds for $i=e-1$. Next, we assume (\ref{eq:Proposition1_Appendix})
has been proved for $j\leq i\leq e$. Note that $s_{t}^{j}(1+\delta_{j-1}S_{t}^{j})$
becomes:
\begin{eqnarray}
s_{t}^{j}(1+\delta_{j-1}S_{t}^{j}) & = & \left(\sum_{u=j}^{e-1}\delta_{u}\prod_{v=j+1}^{u}(1+\delta_{v-1}S_{t}^{v})\right)(1+\delta_{j-1}S_{t}^{j})\\
 & = & \left(\sum_{u=j}^{e-1}\delta_{u}\prod_{v=j}^{u}(1+\delta_{v-1}S_{t}^{v})\right)=s_{t}^{j-1}-\delta_{j-1}\label{eq:InductionSecond}
\end{eqnarray}
With (\ref{eq:InductionSecond}), $P_{t}^{e}s_{t}^{j-1}$ is computed
as:
\begin{equation}
P_{t}^{e}s_{t}^{j-1}=P_{t}^{e}\delta_{j-1}+P_{t}^{e}s_{t}^{j}(1+\delta_{j-1}S_{t}^{j})=P_{t}^{e}\delta_{j-1}+A_{t}^{j}(1+\delta_{j-1}S_{t}^{j})=A_{t}^{j}+\delta_{j-1}P_{t}^{j}=A_{t}^{j-1}\label{eq:InductionThird}
\end{equation}
(\ref{eq:InductionThird}) means (\ref{eq:Proposition1_Appendix})
is shown for $i=j-1$. Thus we have proved (\ref{eq:Proposition1_Appendix})
for $i\leq e-1$.

Now we compute the no-arbitrage drift $\mu_{t}^{i,T_{e}}$. With the
measure-change technique, $\mu_{t}^{i,T_{e}}$ is given as:
\begin{eqnarray}
\mu_{t}^{i,T_{e}} & =- & \frac{\langle d\ln(\frac{A^{a(i)}}{P^{e}}),dW^{(i),T_{e}}\rangle_{t}}{dt}=-\frac{\langle ds^{a(i)},dW^{(i),T_{e}}\rangle_{t}}{dt}\frac{1}{s_{t}^{a(i)}}
\end{eqnarray}
Note that:
\begin{eqnarray}
ds_{t}^{i} & = & d\left(\sum_{u=i}^{e-1}\delta_{u}\prod_{v=i+1}^{u}(1+\delta_{v-1}S_{t}^{v})\right)\\
 & = & \sum_{l=i+1}^{e-1}\sum_{u=l}^{e-1}\left[\delta_{u}\prod_{v=i+1}^{u}(1+\delta_{v-1}S_{t}^{v})\right]\frac{\delta_{l-1}dS_{t}^{l}}{1+\delta_{l-1}S_{t}^{l}}+(...)dt=\sum_{l=i+1}^{e-1}s^{il}\frac{\delta_{l-1}dS_{t}^{l}}{1+\delta_{l-1}S_{t}^{l}}+(...)dt
\end{eqnarray}
Then we obtain $\mu_{t}^{i,T_{e}}$ as:
\begin{eqnarray}
\mu_{t}^{i,T_{e}} & = & -\sum_{u=a(i)+1}^{e-1}\frac{s_{t}^{a(i)u}}{s_{t}^{a(i)}}\frac{\delta_{u-1}\sqrt{\xi_{t}^{u,t}}\rho_{t}^{Y,iu}}{1+\delta_{u-1}S_{t}^{u}},\;1\leq i\leq N_{S}
\end{eqnarray}
 
\end{proof}
\bibliographystyle{unsrt}
\bibliography{SMMBegomi}

\end{document}